\documentstyle[preprint,aps]{revtex}
\draft
\begin{document}
\draft
\title{Oscillatory Tunneling between Quantum Hall
Systems\\}
\author{Tin-Lun Ho}
\address{Physics Department, The Ohio State University, Columbus,
Ohio, 43210 $^{\ast}$ \\
National High Magnetic Field Laboratory, Florida State University, Tallshassee,
FL32306-4005
Physics Department, Hong Kong University of Science and Technology,
Clear Water Bay, Kowloon, Hong Kong\\
Physics Department, Chinese University of Hong Kong,
Shatin, N.T. Hong Kong\\}
\maketitle
\begin{abstract}
Electron tunneling between quantum Hall systems on the same two
dimensional plane separated by a narrow barrier is studied. We show that
in the limit where inelastic scattering time is much longer than the
tunneling time, which can be achieved in practice, electrons
can tunnel back and forth through the barrier continously, leading to an
oscillating current in the absence of external drives. The
oscillatory behavior is dictated by a tunneling gap in the energy
spectrum. We shall discuss ways to generate oscillating currents
and the phenomenon of natural ``dephasing" between the tunneling currents
of edge states.  The noise spectra of these junctions are also
studied. They contain singularites reflecting the existence of
tunneling gaps as well as the inherent oscillation in the system.
\end{abstract}


\narrowtext

\section{Oscillatory Tunneling in Quantum Hall Systems}
\label{sec:level1}

In this paper, we study electron tunneling between quantum Hall (QH)
systems separated by {\em thin} barriers.  Examples of these systems are
shown in figure~\ref{semiclassical} to figure~\ref{capture}.  The
thinness of the barrier allows an electron to tunnel through it many
times before scattered away by inelastic effects.  Oscillatory tunneling
of this kind will occur if the inelastic scattering time $\tau_{in}$ is
much longer than the tunneling time $\tau_{T}$,
\begin{equation}
\tau^{}_{in} >>  \tau^{}_{T}  . \label{osc}
\end{equation}

The existence of oscillatory tunneling can be seen even in the
semiclassical (SC) limit, where electron wave-packets moves in circular
orbits with cyclotron frequency $\omega_{c}$.  When the barrier is
infinite, electrons will undergo a sequence of ``reflected circular
orbits" as shown in fig.\ref{semiclassical}. In the absence of other
scattering mechanisms, electrons having collided with the barrier once
must collide with it again within the cyclotron period. As a result,
they are forever $captured$ by the barrier. (See fig.~{\ref{capture}).
When the barrier is reduced from infinity to a finite value, the
$captured$ electrons on one side of the barrier (say, $L$) can tunnel to
the other side ($R$). Once tunneled, this electron will collide
repeatedly with the barrier and eventually tunnel back to $L$.  When
eq.(\ref{osc}) is satisfied, this back and forth tunneling process can
proceed without interuption, giving rise to an oscillating current in
the absence of external drives.

While the SC picture captures the correct physics, it only tells half
the story.  In a quantum mechanical treatment, we shall see that
different edge states tunnel with different frequencies. Thus, even in
the absence of inelastic scattering, the tunneling current of different
edge states will naturally dephase with each other.  As a result, the
total tunneling current will decrease in time.  However, we show later
that despite dephasing effects, there are ways to generate lasting
current oscillations (thereby reflecting the oscillatory tunneling near
the barrier) without the aid of an a.c. drive.

The crucial question is whether eq.(\ref{osc}) can be achieved. We shall
argue below that this is possible at least for the case of integer
filling.  There are two sources of inelastic scattering: Coulomb
interaction between electrons on the same side of the barrier
(``intra-region" interaction) and that on the different sides
(``inter-region" interaction).  Let us first consider noninteracting
electrons and the limit of infinite barrier. The systems $R$ and $L$ on
both sides of the barrier are now disconnected, reducing to two
semi-infinite systems terminated by a hard wall.  The Landau levels of
such systems are well known, i.e. they bend upward as the barrier is
approached\cite{Halperin}\cite{MacDonald}.  See figure~\ref{inftyspec}
In the presence of intra-region interaction (but without inter-region
interaction), and when the system has integer filling, the edge
electrons will behave like a normal Fermi liquid\cite{Wen}.  The
lifetime $\tau_{in}$ of the quasiparticles will then tend to to infinity
at the Fermi surface, and will dominate over any tunneling time
$\tau_{T}$ introduced by finite barriers.  In other word, eq.(\ref{osc})
can always be satisfied near the Fermi surface when the system has
integer filling, and that electron tunneling near the Fermi surface can
be modelled by that of non-interacting systems.  (Estimates of the
tunneling time is given at the end of the paper).  As we have seen in
fig.~\ref{inftyspec}, the Landau levels of $R$ and $L$ intersect because
they all bend upward near the barrier.  In the presence of tunneling,
these intersections will turn into gaps, (see figure~\ref{general}).  As
we shall see, the unusual features of these junctions are determined by
these gaps.

What is more subtle is the effect of inter-region interactions. While it
is obvious that the tunneling gap can withstand sufficiently weak
inter-region interactions, the situation is less clear for large
inter-region interactions. However, as we show later, it is possible to
map our problem to a solvable model in one dimension (massive Thirring
model). The exact solution of this model shows that the tunneling gap
exists for arbitrary inter-region interaction. Although we have not yet
been able to calculate the current responses for arbitrary inter-region
interaction, the survival of the tunneling gap suggests that the
tunneling characteristics of the non-interacting systems may also survive.

Before proceeding, we stress that the phenomena discussed here requires
thin barriers.  The junction used in many current experiments are
produced by gate voltages and are much smoother than the barriers we
consider here\cite{relation}.
Since magnetic lengths in a 10 Tesla field is about
$80\AA$, and that channels of $100\AA$ wide is feasible in current
technology, the construction of these junctions is possible.  (See also
Section VII for estimates of relevant parameters).

The rest of paper is organized as follows : In Section II, we discuss
the energy spectra in the vicinity of the the barrier for a variety of
external conditions.  In Section III, we derive the effective
Hamiltonian for the tunnel junction as well as the expression of
tunneling current.  In section IV, we suggest ways to generate
oscillatory tunneling currents, and discuss the phenomenon of natural
dephasing.  In section V, we discuss the noise spectrum of the junction,
which reflects directly the existence of tunneling gaps and the inherent
natural oscillations of the system.  In Section VI, we discuss the
effect of inter-region Coulomb interaction.  In Section VII, we give
numerical estimates of various parameters.

\section{The Energy Spectrum Near the Barrier}
\label{sec:level2}

 We have argued in Sec.\ref{sec:level1} that when eq.(\ref{osc}) is
satisfied, tunneling between QH systems with fully filled Landau levels
can be modelled by that of non-interacting electrons.  Although we have
mentioned the general behavior of the spectrum in Section I, we shall
give a detailed desciption here as we shall need it later.
For simplicity, we shall focus on the setup in
fig~\ref{rec}.  The system is periodic in $y$, $\psi(x,y) = \psi(x,y+L)$.
The Hamiltonian in the Landau gauge is
\begin{equation}
H = \frac{1}{2m}p_x^{2} + \frac{1}{2m}\left(p_y
- {e\over c}Bx\right)^{2} +  V(x)
\label{one},
\end{equation}
where $B$ is the external magentic field, and  $V(x)=V_{o}>0$ or $0$
for $|x|>$ or $<a$. (See fig.~\ref{inftyspec}).
The eigenstates
are of the form $\psi_{n,k}(x,y) = L^{-1/2}e^{iky}u_{n,k} (x)$,
$k=(2\pi m)/L$, where $m$ is an integer, and
$u_{n,k}(x)$ is an eigenfunction of
\begin{equation}
H_{k}(x) =  \hbar \omega_{c} \left(
- \frac{1}{2}\ell^{2}\partial_x^{2} +  V_{k}(x) \right) , \,\,\,\,\,\,\,\,
V_k (x) \equiv {1\over 2}(\frac{x}{\ell}-k\ell)^{2} + V(x)/\hbar \omega_{c}
,   \label{three} \end{equation}
with energy $E_{n,k}$. Here,
$\ell = \sqrt{\hbar c/eB}$ is magnetic length and
$\omega_{c} = (eB/mc)$ is the cyclotron frequency.
Eq.(\ref{one}) can be written as
 \begin{equation}  H= \sum_{k}H^{}_{k} \equiv
\sum_{k} \sum_{n=0}^{\infty}E^{}_{n,k} a^{+}_{n,k}a_{n,k}, \label{hentire}
\end{equation}
where $a_{n,k}$ is annihilation operator of $\psi^{}_{n,k}$.
The existence of oscillatory tunneling  near the barrier
can be seen from the fact that $V_{k}(x)$
reduces to  a degenerate double well
as $k\rightarrow 0$. It is well known that when an electron is placed in one
side of the double well,
it will tunnel back and forth between the wells with a
 frequency  given by the excitation energy from the
ground state to the first excited state.

Although both $u_{n,k}(x,y)$ and $E_{n,k}$ can be obtained by analytic
methods\cite{MacDonald},
they can be easily understood in the limit of high barriers.
When $V_{o}=\infty$, $L$ and $R$ become two disconnected semi-infinite
systems,
 $H \rightarrow H_{L} + H_{R}$, \begin{equation}
H_{L} = \sum_{k=0}^{\infty}\sum_{n=0}^{\infty}
\epsilon^{L}_{n,k} c^{+}_{n,k}c_{n,k}, \,\,\,\,\,\,
H_{R} = \sum_{k=0}^{\infty}\sum_{n=0}^{\infty}
\epsilon^{R}_{n,k} d^{+}_{n,k}d_{n,k},   \label{hinfinite}
\end{equation}
where $\epsilon^{L}_{n,k}$ and $\epsilon^{R}_{n,k}$ are the Landau levels
of   $L$ and $R$ in the limit $V_{o}=\infty$.
$c_{n,k}$ and $d_{n,k}$ are the corresponding eigenstates.
The behavior of the Landau levels $\epsilon^{L}_{n,k}$ and
 $\epsilon^{R}_{n,k}$ as a function of $k$ have been studied by a
 number of authors\cite{Halperin}\cite{MacDonald}. (See also
fig.~\ref{inftyspec} and fig.~\ref{general}).
In the bulk of $L$, $\ell k<<-1$,
$\epsilon^{L}_{n,k}=(n+1/2)\hbar\omega_{c}$. $\epsilon^{L}_{n,k}$
begins to deviate appreciably from
its bulk value about a magnetic length away from the wall, $\ell k \alt -1$.
The entire curve increases
monotonically (to infinity) as $k$ increases, passing
through $(2n+3/2)\hbar\omega_{c}$ at the barrier,
(i.e. when $\ell k= -a/\ell$). $\epsilon^{R}_{n,k}$ has an identical
behavior in the reverse $k$ direction.
When $V_{o}$
 is reduced from infinity to a finite value, the
intersections of the spectra of $L$ and $R$ will turn into ``tunneling" gaps.
The two sets of energy curves $\{ \epsilon^{L}_{m,k} \}$
and $\{ \epsilon^{R}_{m',k} \}$ now turn into a single set
$\{ E_{n,k} \}$, which we shall refer to as the $n$-th Landau level of the
$entire$ system.
Each curve $E_{n,k}$ is a smooth function in $k$. It reduces to
$\epsilon^{L}_{n,k}$
 and  $\epsilon^{R}_{n,k}$ for $\ell k<< -1$ and $\ell k >> 1$.

The qualitative features of the wavefunctions
$u_{n,k}$ can be determined from the effective
potential $V_{k}(x)$. (See figure~\ref{figwf}).  If $\phi^{L}_{n,k}(x)$ and
$\phi^{R}_{n,k}(x)$  are the eigenstates of $L$ and $R$
in the infinite barrier
limit, (hence $\phi^{L}_{n,k}(x)=0$ for $x>0$, and
$\phi^{R}_{n,k}(x)=0$ for $x<0$),
then for high (but finite) barriers, we have (see fig.~\ref{figwf})

\begin{equation}
u^{}_{0(1),k} \approx  \left( \phi^{L}_{0,k}(x) +(-)
\phi^{R}_{0,k}(x) \right)/\sqrt{2}  \,\,\,\,\,\, {\rm for} \,\,\,\,
-1<<\ell k <<1 ,     \label{phidef}
\end{equation}
\begin{equation}
u_{0(1),k}(x) \approx \phi^{L}_{0(1),k}(x), \,\,\,\,\,\,
{\rm for}\,\,\,\, \ell k << -1 ;    \end{equation}
\begin{equation}
u_{0(1),k}(x) \approx  \pm \phi^{R}_{0(1),k}(x), \,\,\,\,\,\,
{\rm for}\,\,\,\,  \ell k >> 1 .    \end{equation}

In the absence of voltage bias between $L$ and $R$, it can be seen
 from fig.~\ref{general} that the lowest tunneling gap
of $L$ and $R$ (which occurs at $k=0$) lie above
the first ``bulk" Landau level, i.e. $(3/2)\hbar\omega_{c}$.
The location of the tunneling gap, however, can be easily changed
 by applying a voltage bias. (See fig.~\ref{figbias}).
 Note that in the presence of a voltage bias $V$,
 $H$ is still diagonal in $k$ and is still given by eq.(\ref{hinfinite})
 except that the spectrum $E_{n,k}$ and the eigenstates $a_{n,k}$
 now functions of $V$.

When the spins of the electrons are taken into account, the spectrum of
$L$ and $R$ in the infinite barrier limit consists of two sets
of Landau levels differing from each other by the
Zeeman energy. Since $V(x)$ does not flip spins, the intersections of the
opposite spin Landau levels will not turn into gaps when
$V_{o}$ becomes finite.

To conclude this section, we derive the expression for the current in the
$x$-direction. If we
define the number of particle to the left and to the right of the barrier as
\begin{equation}
N_{L}(t) = \int^{0}_{-\infty} dx \hat{\psi}^{+}(x,y;t)\hat{\psi}(x,y;t),
\,\,\,\,\,\,\,
N_{R}(t) = \int^{\infty}_{0} dx \hat{\psi}^{+}(x,y;t)\hat{\psi}(x,y;t)
\end{equation}
the current in $x$ is then $I(t) =e\dot{N_{L}}
= - e\dot{N_{R}}$.  Using the fact that $\hat{\psi}(x,y,t) = $ \\
$\sum^{}_{n,k}\left[L^{-1/2}e^{iky}
u^{}_{n,k}(x)\right] a^{}_{n,k}e^{-iE_{n,k}t/\hbar}$, we can write
\begin{equation}
I(t) = \sum_{k}I_{k}(t) = \sum_{k}\frac{ie}{\hbar}\sum_{n,m}
[E_{n,k} - E_{m,k}] g^{}_{n,m}(k)
a^{+}_{n,k}a^{}_{m,k}e^{i[E_{n,k}-E_{m,k}]t/\hbar},
\label{current}
\end{equation}
\begin{equation}
g_{n,m}(k) = \int^{0}_{-\infty} u_{n,k}(x) u_{m,k}(x) dx .
\end{equation}
Note that only terms with $n\neq m$  contribute to the current as
$g_{n,m}(k)$ reduces to the overlap of two orthogonal states
in $L$ or $R$ in the $V_{o}=\infty$ limit,
\begin{equation}
g_{n,m}(k) \approx 0  \,\,\,\,\,\,
{\rm for} \,\,\,\,\ell |k|>1 . \label{glargek}
\end{equation}
For this reason, {\em  we can from now on foucs on
the range $\ell |k| \leq 1$ in eq.(\ref{current}). }

Limiting to the lowest two Landau levels, eq.(\ref{current})
becomes  \begin{equation}
I(t) =  \sum_{k}\frac{ie}{\hbar} T^{}_{k}
a^{+}_{1,k}a^{}_{0,k}e^{i[E_{1,k}-E_{0,k}]t/\hbar} + h.c. ,
\,\,\,\,\,\,\,\,\,\,\, T^{}_{k} = [E_{1,k} - E_{0,k}] g^{}_{1,0}(k)
.   \label{newcurrent}
\end{equation}
For $\ell |k| \leq 1$, eq.(\ref{phidef}) implies
\begin{eqnarray}
g_{1,0}(k) & \approx & \frac{1}{2} \int^{0}_{-\infty} dx
\left[ |\phi^{L}_{0,k}(x)|^{2}
- |\phi^{R}_{0,k}(x)|^{2} \right] \\
 & = & \frac{1}{2} \int^{0}_{-\infty}dx  |\phi^{L}_{0,k}(x)|^{2}
 =\frac{1}{2} \,\,\, . \label{gsmallk}
 \end{eqnarray}
With eq.(\ref{gsmallk}) and eq.(\ref{glargek}), we have
\begin{eqnarray}
T_{k} & \approx  (E_{1,0} - E_{0,0})/2 \equiv \Delta_{o}/2
& \,\,\,\,\, {\rm for} \,\,\, \ell |k| << 1 \label{ok} \\
 & \approx 0  & \,\,\,\,\, {\rm for} \,\,\, \ell |k| >1.  \label{ook}
 \end{eqnarray}

\section{Effective Tunneling Hamiltonian and The tunneling Current}
\label{sec:level3}

In this section and the next two, we shall focus on
the tunneling between the lowest Landau level of $L$ and
$R$. For simplicity. we shall also consider
the case of zero bias.
The results derived here can be generalized easily to other Landau
levels and to non-zero bias.
The Hamiltonian of the entire system, eq.(\ref{hentire}),
now reduces to
\begin{equation}
H = \sum_{k} \left( E^{}_{0,k}a^{+}_{0,k}a^{}_{0,k} +
E^{}_{1,k}a^{+}_{1,k}a^{}_{1,k}\right). \label{reducedh} \end{equation}
As discussed in Sec.I, only those $k$'s in the range  $\ell |k|\leq 1$
contribute to the current eq.(\ref{current}). Within this range,
  $E_{0,k}$ and $E_{1,k}$  are close to
$\epsilon^{L}_{0,k}$
and $\epsilon^{R}_{0,k}$ except at $k=0$,
(i.e. the intersection of $\epsilon^{L}_{0,k}$ and
$\epsilon^{R}_{0,k}$),  where a gap $\Delta_{o}$ is opened up.
(See also fig.~\ref{general} and
fig.~\ref{before}).
For later discussions, we define
\begin{equation}
E_{1,k} - E_{0,k} \equiv E_{k}  , \,\,\,\, \Delta_{o}
\equiv E_{k=0}=E_{1,0}-E_{0,0} .
 \label{defek}   \end{equation}
The tunneling phenomenon contained in eq.(\ref{reducedh}) is more transparent
if $H$ is written in the form of a tunneling Hamiltonian. Defining energies
$\epsilon_{L,k}$, $\epsilon_{R,k}$, and tunneling matrix element $T_{k}$ as
\begin{equation}
\epsilon^{}_{L,k} - \epsilon^{}_{R,k} =
\epsilon^{L}_{0,k} - \epsilon^{R}_{0,k}
    \equiv   \epsilon_{k} , \label{addcon} \end{equation}
\begin{equation}
\epsilon_{L(R),k}  =  \epsilon^{L(R)}_{0,k} + \zeta_{k}, \,\,\,\,\,\,\,
\zeta_{k} = \frac{1}{2}\left[ E_{1,k}+E_{0,k} -\epsilon^{L}_{0,k}
-\epsilon^{R}_{0,k} \right],   \end{equation}
\begin{equation}
T_{k} = \frac{1}{2}\sqrt{ E_{k}^{2}  - \epsilon_{k}^{2} }, \label{tdef}
\end{equation}
eq.(\ref{reducedh}) can be written as
 \begin{equation}
H = H_{o} + H_{T} = \sum_{k} \left( \epsilon^{}_{L,k}c^{+}_{L,k}c^{}_{L,k}
+ \epsilon^{}_{R,k}c^{+}_{R,k}c^{}_{R,k}\right) -
  \sum_{k}\left( T^{}_{k}c^{+}_{L,k}c^{}_{R,k} + h.c. \right),  \label{htunn}
\end{equation}
where
\begin{equation}
\left( \begin{array}{c}
c_{L} \\c_{R} \end{array} \right)_{k} = \hat{U}_{k} \left( \begin{array}{c}
a_{o} \\a_{1} \end{array} \right)_{k},  \,\,\,\,\,\,
 \hat{U}_{k} = \left( \begin{array}{cc}
v & u \\ u & -v \end{array} \right)_{k},
\label{clcr} \end{equation}
\begin{equation}
u_{k} =  \sqrt{ \frac{1}{2}\left(1 +
 \frac{\epsilon_{k}}{E_{k}}\right) }, \,\,\,\,\,\,
v_{k} =  \sqrt{ \frac{1}{2}\left(1 -
\frac{\epsilon_{k}}{E_{k}}\right) } .   \label{uvdef}
\end{equation}
The phases of $c_{L,k}$ and $c_{R,k}$ have been chosen so that
$u_{k}, v_{k}$, and $T_{k}$ are all real. [The relation between $T_{k}$ defined
in eq.(\ref{tdef}) and that in
eq.(\ref{newcurrent}) will be clear shortly].
Eq.(\ref{uvdef}) also implies that
\begin{equation}
T_{k} = u_{k}v_{k}E_{k}.  \label{newtk}
\end{equation}

Although strictly speaking $\zeta_{k}\neq 0$, it can be taken as zero as it is
much smaller than
$\epsilon^{L(R)}_{0,k}$.
As a result, $\epsilon_{L,k}$,
$\epsilon_{R,k}$, $c_{L,k}$, and $c_{R,k}$ are well aproximated by
$\epsilon^{L}_{0,k}$,
$\epsilon^{R}_{0,k}$,
$c_{0,k}$, and $d_{0,k}$,  (see eq.(\ref{hinfinite})), even though they are not
exactly the same. $H_{o}$ in eq.(\ref{htunn}) can therefore be interpreted as
the
Hamiltonian of $L$ and $R$ in the infinite barrier limit, and $H_{T}$ describes
the tunneling between them.

[There is another point worth noting. In the conventional tunneling
Hamiltonian,
the tunneling term $H_{T}$ is usually written as
$\sum_{k,k'}(T_{k,k'}c^{+}_{L,k}c_{R,k'} + h.c.)$,
 whereas in eq.(\ref{htunn}) $k$ is conserved during tunneling processes.
This is entirely a consequence
of the symmetry of the systems in fig.~\ref{ciredge} and
fig.~\ref{rec}].

Next, we turn to the tunneling current.
Defining the number of particles to the left and to the right
as $N_{L} = \sum_{k}c^{+}_{L,k}c_{L,k}$, and $N_{R} =
\sum_{k}c^{+}_{R,k}c_{R,k}$,
the current in $x$ is then  $I(t) = e\dot{N}_{L} = -e\dot{N}_{R}$, or
explicitly,
\begin{equation}
 I(t) = \frac{ie}{\hbar} \sum_{k} \left(T^{}_{k} c^{+}_{L,k}(t)c_{R,k}(t)
 - h.c. \right) .    \label{currentori}
\end{equation}
Using eq.(\ref{uvdef}) and the fact that $a_{0(1),k}(t) =
a_{0(1),k}e^{-iE_{0(1),k}t/\hbar}$, we can write eq.(\ref{currentori}) as
\begin{equation}
 I(t)  = \frac{ie}{\hbar} \sum_{k}T_{k}\left( a^{+}_{1,k}a^{}_{0,k}
e^{iE_{k}t/\hbar} +h.c. \right).  \label{fcurrent1} \end{equation}
Using eq.(\ref{uvdef}) again, we can rewrite \begin{equation}
 I(t) = \frac{e}{\hbar}\sum_{k}
 \left[ \nu_{k}(t)\left( c^{+}_{R,k}c^{}_{R,k}
-c^{+}_{L,k}c^{}_{L,k}\right) + \eta_{k}(t)\left(c^{+}_{L,k}c^{}_{R,k}
+h.c.\right) \right],   \label{fcurrent2}  \end{equation}
where $\nu_{k}$ and $\eta_{k}$ are defined as
\begin{equation}
\nu_{k}(t) = 2(T_{k}u_{k}v_{k}){\rm sin}(E_{k}t/\hbar), \,\,\,\,\,\,
\eta_{k}(t) = iT_{k} \left( e^{iE_{k}t/\hbar}u_{k}^{2}
+ e^{-iE_{k}t/\hbar}v_{k}^{2} \right).  \label{nutau} \end{equation}
Comparing  eqs.(\ref{fcurrent1}) and (\ref{newtk}) with
eq.(\ref{newcurrent}), one notes that these  two definitions of $T_{k}$ are
consistent if $g_{0,1}$ in eq.(\ref{newcurrent})
is identified as  $u_{k}v_{k}$ in eq.(\ref{newtk}).
{}From eq.(\ref{uvdef}), one can also see that the asympotic form
eq.(\ref{ok}) is also satisfied by both definitions.

The expressions eqs.(\ref{fcurrent1}) and (\ref{fcurrent2})
represent the major difference
 between oscillatory
tunneling and the usual type of electron tunneling (such as those in
normal and Josephson junctions), where first order perturbation theory in
$H_{T}$ provides an adequate description of the tunneling current,
\begin{equation}
I(t) - I(0) = \frac{i}{\hbar}\int^{t}_{-\infty}
[\tilde{I}(t), \tilde{H}_{T}(t')]dt' , \,\,\,\,\,\,\,\,
\tilde{A}(t) \equiv e^{iH_{o}t/\hbar} A e^{-iH_{o}t/\hbar}.
\label{oldcurrent}
\end{equation}
In the conventional treatment, eq.(\ref{oldcurrent}), higher order terms
corresponding to multitunneling processes
are ignored. In contrast,  the time dependences in
eq.(\ref{currentori})-(\ref{fcurrent2}) are generated
by the full Hamiltonian $H$, which is
amount to extending the perturbation series eq.(\ref{oldcurrent})
to infinite order.  (Note that the infinite series is necessary to generate a
gap in the
spectrum).

The reason that the perturbative result eq.(\ref{oldcurrent}) is
applicable for most systems is because
multitunneling processes are usually subpressed by quantum diffusion
even in the absence of inelastic scattering.
In the usual case, tunneling takes place between electronic
states that are extended over the bulk of the sample ($L$ or $R$).
 Once tunneled across, the
electron leaves the barrier on the time scale of quantum diffusion.
The time to travel a distance comparable to the barrier width
$a/v_{F}$,
which is usually much shorter than the tunneling time. As a result, the
tunneling current can be accounted for by first order
perturbation theory.  However, for the junctions we consider, tunneling takes
place
between edge states which are localized near the barrier.
The electron has no where to go after tunneling but to tunnel back. The
continuous back and forth tunneling renders the conventional scheme
eq.(\ref{oldcurrent}) inadequate.

{\em A Simple Model} : Near the intersection point, one can linearize the
infinite barrier spectrum $\epsilon_{k}$ such that
\begin{equation}
\epsilon_{k} = v^{}_{F}\hbar k  \label{linear} \end{equation}
where $v_{F}$ is the Fermi velocity which  is of the order of
$\ell \omega_{c}$.    When the tunneling is weak,
$\Delta_{o} <<\hbar\omega_{c}$, the region in $k$-space where $E_{k}$ differs
significantly from $\epsilon_{k}$ is
$\ell |k| \leq \frac{\Delta_{o}}{\hbar \omega_{c}}$.
We can therefore model $E_{k}$ as
 \begin{equation}
E^{2}_{k} = \Delta_{o}^{2}  + \epsilon_{k}^{2}\,\,\,\,.   \label{approxispec}
\end{equation}
In terms of this model,
eq.(\ref{tdef}) and eq.(\ref{newtk}) become
\begin{equation}
T^{}_{k} = \frac{1}{2}\Delta^{}_{o}, \,\,\,\,\,\,\, u^{}_{k}v^{}_{k}
= \frac{\Delta^{}_{o}}{2E^{}_{k}} .   \label{modeltk} \end{equation}

\section{Oscillations of the Tunneling Current}
\label{sec:level4}

{}From eqs.(\ref{fcurrent1}) and (\ref{fcurrent2}), we see that the tunneling
current  is made up of different edge state components
$I_{k}$, each of which oscillates at a different frequency
$\omega_{k} = E_{k}/\hbar$. In this section, we   discuss ways to
generate natural current
oscillations, and to discuss the dephasing between different current
components. For simplicity, let  both
$L$ and $R$ have identical chemcical potentials
(i.e. $\mu_{L}=\mu_{R}=\mu$), and that $\mu$ is below the tunneling gap.
(See fig.~\ref{before}).
The corresponding Fermi vectors in $L$ and $R$ are $-k_{F}$ and
$k_{F}$  respectively. The quantum state of the
 system is then  $|\Psi> = \prod_{k<-k_{F}} c^{+}_{L,k}
\prod_{p>k_{F}} c^{+}_{R,p}|0>$. [That we take  the initial state as $|\Psi>$
instead of the true ground state of the entire system
$|\Psi_{o}> = \prod_{|k|>k_{F}}a^{+}_{0,k}$ is because the relaxation from
$|\Psi>$ to $|\Psi_{o}>$ requires inelastic processes, which are ineffective
when eq.(\ref{osc}) is satisfied.]
The tunneling  current  is
\begin{equation}
<I(t)> =  \frac{e}{\hbar}\sum_{k} \nu_{k}(t)\left( <c^{+}_{R,k}c^{}_{R,k}>
-<c^{+}_{L,k}c^{}_{L,k}> \right),   \label{currentave}
\end{equation}
where the average is with respect to $|\Psi>$. It is clear that $|\Psi>$
will not generate any current as the current components in $L$ and $R$
cancel each other, ($\nu_{k}(t) = \nu_{-k}(t)$, hence)
$<I(t)> = \frac{e}{c}\left( \sum_{k< -k_{F}} -
\sum_{k>k_{F}}\right)\nu_{k}(t)=0$.

The simplest way to generate a single (of a small number of)
oscillating current component is to move all the edge states $k$ to the
right by a small amount, i.e. shifting $k$ to ($k + \theta/L$).  This
shift amounts to changing the periodic boundary condition of the
wavefunction to
$\psi(x,y) = e^{i\theta} \psi(x,y+L)$. Returning to the cylindrical geometry
fig.~\ref{ciredge}, this change of boundary condition corresponds to
passing a fraction
($\frac{\theta}{2\pi}$) of a flux quantum through the center hole.
When half of a flux
quantum is passed,  ($\theta = \pi$), we have in effect added an electron
on top of the Fermi sea in $L$.
(See fig.~\ref{before}).  The tunneling current is therefore
\begin{equation}
<I(t)> =  -\nu_{k_{F}+\pi/L} \approx -\nu_{k_{F}} =
  -\frac{e}{\hbar} 2(Tuv)_{k_{F}} {\rm sin}
\left(E_{k_{F}}t/\hbar\right) . \label{singlecur}
\end{equation}
For chemical potentials slightly below the tunneling gap,
$\ell k_{F}<<1$, eq.(\ref{newtk}) implies
$(Tuv)_{k_{F}}=2E_{k_{F}}(uv)_{k_{F}}^{2}=E_{k_{F}}$. We can then write
eq.(\ref{singlecur}) in a very simple form
\begin{equation}
<I(t)> =  - e \left( E_{k_{F}}/\hbar\right) {\rm sin}
\left(E_{k_{F}}t/\hbar\right).    \label{currentsingle}
\end{equation}

If, instead of pushing a flux quantum through $L$, we introduce a
chemical potential difference between $L$ and $R$ at time $t=0$,
[ ($\mu_{L} = \mu_{R}) \rightarrow
(\mu_{L} = \mu + eV/2, \mu_{R} = \mu - eV/2$)]. The
Fermi wavevectors in $L$ and $R$ are then changed to
$-k_{F} + \delta k_{F}$ and $k_{F} + \delta k_{F}$, (see fig.~\ref{figdeltav}),
\begin{equation}
\delta k_{F} = eV/(\hbar v_{F}).   \label{deltakf}
\end{equation}
The quantum state in eq.(\ref{currentave})   now becomes
$|\Psi>=(\prod_{k<-k_{F}+\delta k_{F}} c^{+}_{L,k})
(\prod_{p>k_{F}+\delta k_{F}} c^{+}_{R,p})|0>$. The current at $t>0$ is then
\begin{equation}
<I(t)> = - \frac{e}{\hbar}
\sum_{[k]} \nu_{k}(t) ,  \label{currentnu}
\end{equation}
where $[k]$ denotes the range of exicted edge states
$|k+k_{F}|\leq \delta k_{F}$.
Each of the $\nu_{k}$ term oscillates with frequency $E_{k}/\hbar$.
If the entire range $[k]$ lies in the linear region of the spectrum,
(hence $E_{k} \approx \epsilon_{k}=\hbar v_{F} k$),
(see fig.~\ref{figdeltav}), then the states at the opposite end of the interval
will be
the first ones to be out of dephase with each other,  as they have maximum
frequency difference.
This takes place at time $\tau^{(i)}_{dp} = \hbar/(E'_{k_{F}}\delta k_{F})
\approx
\hbar/(eV/2)$, referred to as the ``initial" dephasing time. As time increases,
the coherence of the states in $[k]$ reduces as more and more
states at different
ends of the interval keep dephasing
with each other.
(See fig.\ref{figdeltav}). When $t \sim \tau^{(f)}_{dp}=
\hbar/[E_{k_{F}}'(2\pi/L)]
= L/[2\pi v_{F}]$, referred to as the ``final" dephasing time,
only one or two states in the vicinity of $k_{F}$ remain coherent.

During the dephasing period, $\tau^{(i)}_{dp} < t < \tau^{(f)}_{dp}$,
the summand in eq.(\ref{currentnu}) is
sufficiently smooth that the sum can be approximated by the integral
\begin{equation}
 <I(t)> = - \frac{e}{\hbar}\frac{L}{2\pi}
\int^{-k_{F}+\delta k_{F}}_{-k_{F}-\delta k_{F}}
  (2T_{k}u_{k}v_{k})
 {\rm sin}\left( \frac{E_{k}t}{\hbar}\right) dk .
\label{currentdv}\end{equation}
Expanding the integrand about $k_{F}$, eq.(\ref{currentdv}) becomes
\begin{equation}
  <I(t)> = - \frac{e}{\hbar}\frac{L}{2\pi}
(2Tuv)_{k_{F}}
{\rm sin}\left(E_{k_{F}}t/\hbar\right)
\frac{ 2{\rm sin}\left( [E'_{k_{F}}\delta k_{F} t]/\hbar\right) }
{E'_{k_{F}} t/\hbar}  +  O(t^{-2}) + ... ,  \label{currentok} \end{equation}
Initially, (for $t\approx 0$), eq.(\ref{currentdv}) gives
 \begin{equation}
<I(t)> \approx - \frac{e}{\hbar} \left(L\delta k_{F}/2\pi\right)
(2Tuv)^{}_{k_{F}}{\rm sin}\left(E^{}_{k_{F}}t/\hbar\right) ,
\label{currentgp} \end{equation}
which is the single electron current eq.(\ref{currentsingle}) multiplied
by  the number of electrons participating in tunneling,
$\left( L\delta k_{F}/2\pi \right)$.  Dephasing effect
 causes this current to decrease as $1/t$, (see eq.(\ref{currentok}) ).
At time $t\approx \tau^{(f)}_{dp}$, most of the terms in
eq.(\ref{currentnu}) have undergone many oscillations except for a few terms
near $k=0$. The magnitude of the current is then reduced to that
comparable to a single electron, eq.(\ref{currentsingle}).

Let us consider a different situation where initially $k_{F} = 0$.
The range $[k]$ is then symmetric about $k=0$. (See fig.~\ref{figdeltav}).
The tunneling
current eq.(\ref{currentnu}) becomes
\begin{equation}
<I(t)> = \frac{2e}{\hbar} \sum_{0\leq k < \delta k_{F}} \nu_{k}(t) ,
\label{currentqu}
\end{equation}
The largest frequency difference among different $k$ terms is still
$eV/\hbar$, whereas the minimum frequency difference becomes
$\delta \omega_{k=0} = (E_{2\pi/L} - E_{0})/\hbar = \frac{1}{2}
\left(\hbar^{2} v_{F}^{2}/\Delta_{o}\right)(2\pi /L)^{2}/\hbar$.
Therefore, we still have  $\tau^{(i)}_{dp}= \hbar/(eV)$, while
the final dephasing time becomes $\tau^{(f)}_{dp}= 2\pi /\delta \omega_{k=0}$.
For $\tau^{(i)}_{dp} > t > \tau^{(f)}_{dp}$, eq.(\ref{currentqu})
can be written as
 \begin{eqnarray}
 <I(t)> & = & - \frac{e}{\hbar}\frac{L}{2\pi}
(2Tuv)^{}_{k=0} \int^{\delta k_{F}}_{-\delta k_{F}}  dk \,\,
{\rm sin}\left(
[\Delta_{o} + \frac{1}{2}E''_{k=0}k^{2}]t
/\hbar \right)  +  O(...)  \\
 & = & - \frac{e}{\hbar}\left( \frac{L\delta k_{F}}{2\pi}\right)
 \left[ 2 \Delta_{o} {\rm sin}\left( \Delta_{o}t/\hbar \right) \right]
\sqrt{  \frac{\pi \hbar \Delta_{o}}{(eV)^{2} t} }
C\left[\sqrt{  \frac{(eV)^{2} t}{\pi \hbar \Delta_{o}} } \right] ,
\label{cintegral}
\end{eqnarray}
where ${\rm C}(x) \equiv \int^{x}_{0} {\rm cos}(u^{2}) du$ is the Fresnel
integral  which  approaches $1/2$ as $x \rightarrow \infty$.
In deriving eq.(\ref{cintegral}), we have made use of eq.(\ref{deltakf}).

{}From eq.(\ref{cintegral}), we can see that as the chemical potential
$\mu$ sweeps through the gap, the dephasing processes slows down,
changing from $t^{-1}$ to $t^{-1/2}$ for large $t$.  The final dephasing
time $\tau^{(f)}_{dp} = \hbar/\delta \omega_{k=0}= (L/2\pi v_{F})
(\Delta_{o}/[\hbar
v_{F}(2\pi/L)])$ is much longer than that in the
previous case, $(L/2\pi v_{F})$,  as the factor $\Delta_{o}/[\hbar
v_{F}(2\pi/L)]$  is typically much larger than 1. (See also Section VII).

\section{Noise Spectrum}
\label{sec:label5}

The oscillatory tunneling of the edge states can also be detected through
the noise spectrum,
$S(\omega) = \int^{\infty}_{-\infty} S(t) e^{i\omega t} dt$,
$S(t) = \frac{1}{2}< [I(t),I(0)]_{+}>$.  When both $L$ and $R$
have identical chemical potentials $\mu$,
  eq.(\ref{fcurrent2}) and eq.(\ref{nutau}) imply that
\begin{equation}
S(t) =
\left(\frac{e}{\hbar}\right)^{2} \sum_{k}{'} {\rm Re}[ \eta^{\ast}_{k}(t)
\eta_{k}(0)]
=  \left(\frac{e}{\hbar}\right)^{2} \sum_{k}{'} |T_{k}|^{2}
 {\rm cos}\left(\frac{E_{k}t}{\hbar}\right)
 \end{equation}
\begin{equation}
\sum_{k}{'}(\cdot \cdot \cdot) \equiv \sum_{k} (\cdot \cdot \cdot)
\left( f(\epsilon^{}_{L,k})\bar{f}(\epsilon^{}_{R,k})
+ f(\epsilon^{}_{R,k})\bar{f}(\epsilon^{}_{L,k})  \right)
\end{equation}
where $f(x) = (e^{(x-\mu)/k_{B}T} +1 )^{-1}$ is the Fermi function, $T$
is the temperature, and $\mu$ is the chemical potential.
The noise spectrum is
\begin{equation}
S(\omega) = \left(\frac{e}{\hbar}\right)^{2} \sum_{k}{'} |T_{k}|^{2} \pi
\left[ \delta (\omega - E_{k}/\hbar) +
\delta (\omega + E_{k}/\hbar) \right]
\end{equation}
At $T=0$,  we have $\sum'_{k} \rightarrow
(L/2\pi)\left( \int^{-k_{F}}_{-\infty} + \int^{\infty}_{k_{F}}\right)$.
For $\omega >0$, we have
\begin{eqnarray}
S(\omega) & = &
\left(\frac{e}{\hbar}\right)^{2} L \int^{\infty}_{k_{F}}
|T_{k}|^{2}  \delta (\omega - E_{k}/\hbar) dk       \\
 & = & \left(\frac{e}{\hbar}\right)^{2} L \hbar
\left(|T_{k}|^{2}  \frac{dk}{dE_{k}} \right)_{\hbar \omega = E_{k}}
  \,\,\,\,\,\,    {\rm for} \,\, \hbar \omega > E^{}_{k_{F}}  \label{s1} \\
  & = & 0   \,\,\,\,\,\,\,\,\,\,\,\,\,\
  {\rm for} \,\, \hbar \omega < E^{}_{k_{F}}.  \label{s2}
\end{eqnarray}

Using the simple model at the end of Sec.III, eq.(\ref{s1}) and
eq.(\ref{s2}) becomes
\begin{eqnarray}
S(\omega) & = &  \left(\frac{e}{\hbar}\right)^{2}
\frac{L\Delta_{o}^{2}}{v_{F}}
\frac{\omega}{  \sqrt{\omega^{2} - (\Delta_{o}/\hbar)^{2}}  }
\,\,\,\,\,\,   {\rm for}\,\,\,\,\, \hbar \omega \leq E_{k_{F}}  \\
  & = & 0 \,\,\,\,\,\,\,\,\,\,\,\,\, {\rm otherwise}.
  \end{eqnarray}
Note that $k_{F}$ (hence $E_{k_{F}}$) depends on $\mu$. When $\mu$ lies
outside the gap, $k_{F}\leq 0$, and $S(\omega)$ shows a cusp at
$\omega = E_{k_{F}}/\hbar$. When $\mu$ lies inside the gap, $S(\omega)$
shows a square root divergence at $\omega = \Delta_{o}/\hbar$.
(See fig.~\ref{fignoise})

The noise in the tunneling current will generate a similar noise spectrum
$S_{H}(\omega)$
in the Hall current $I_{H}$. The Hall current in $L$ is
$I_{H}=L^{-1}(e/\hbar)\sum_{k}(\partial \epsilon^{L}_{k}/\partial k)
c^{+}_{L,k}c^{}_{L,k}$, and $S_{H}$ is defined as
$S_{H}(t) = \frac{1}{2}<[\delta I_{H}(t), \delta I_{H}]_{+}>$,
where $\delta I_{H} = I_{H} - <I_{H}>$. Using eq.(\ref{clcr}),
it is straightforward to work out this noise spectrum,
\begin{equation}
S_{H}(\omega) = \left(\frac{e}{2\hbar L}\right)^{2}\left(\frac{L}{2\pi}\right)
\int^{\infty}_{k_{F}} \left(\frac{\partial \epsilon_{k}}{\partial k}\right)^{2}
(u_{k}v_{k})^{2} \pi \delta (\omega -E_{k}/\hbar) .
\label{hallcurnoise}
\end{equation}
where we have used the fact that $2\epsilon_{L,k} = \epsilon_{k}$ for unbiased
junctions.  Using eq.(\ref{newtk}) and eq.(\ref{linear}), we have
\begin{equation}
S_{H}(\omega) = \frac{1}{8}\left(\frac{v_{F}}{L\omega}\right)^{2}S(\omega).
\label{noiserelation}
\end{equation}
The noise spectrum of the Hall current is proportional to that of the tunneling
current. While $S_{H}(\omega)$ may be difficult to measure in geometries like
fig.~\ref{rec}, it is easy to measure in the junctions shown in
fig.~\ref{recring}
by measuring the noise spectrum of the Hall voltage,
which is simply $(\hbar/e)^{2}S_{H}(\omega)$.
Even though the junctions in fig.~\ref{recring} and fig.~\ref{rec} are not the
same, the physicis of oscillatory tunneling are identical in both cases.
The nosie spectrum of the tunneling current in fig.~\ref{recring} should have a
divergence as any self oscillating system does, which should show up in the
noise  spectrum of the Hall voltage.

\section{Interactions between edge states on different sides of the barrier}
\label{sec:level6}

So far, we have ignored interactions between edge states on different
sides of the barrier. When these interactions are included, the effecitve
Hamiltonain eq.(\ref{htunn}) becomes
\begin{equation}
H=H_{o} + H_{T} + H_{int}, \,\,\,\,\,\,\,
H_{int} = \epsilon^{-1} \sum_{q}U(q) \rho^{}_{L}(q)\rho^{}_{R}(-q),
\label{thirring}
\end{equation}
where $\rho^{}_{L(R)}(q) = \sum^{}_{K}c^{+}_{k+q, L(R)}c^{}_{k,L(R)}$,
$U(q) = \int^{L/2}_{-L/2} e^{-iqy}(4a^{2} + y^{2})^{-1/2}dy$, and
$\epsilon$ is a dielectric constant.     Eq.(\ref{thirring}) is precisely the
massive Thirring model and its spectrum can be solved exactly
by the Bethe Ansatz\cite{Baxter}.
It is known from the exact solution that there is always a gap in the
spectrum for all $U$.
Since the singularities in the noise spectrum and the
minimum frequency of the oscillatory tunneling current are due to the
existence of the tunneling gap, we expect these features will persist
in the presence of interaction effects.

\section{Estimates of the key parameters}
\label{sec:level7}

Numerical estimates for the parameters in Sec.IV
are given in Table 1.  In these estimates, we take
$m = 0.067m_{e}$, where $m_{e}$ is the mass of the electron.  The barrier
height has been taken as 1 ev.
The tunneling gap $\Delta_{o}$ is calculated by the quasiclassical
method\cite{Landau}, and is given by
\begin{equation}
\Delta_{o}  = \frac{\hbar \Omega}{\pi}
e^{-\frac{1}{\hbar}\int^{a}_{-a}|p| dx} \end{equation}
where $2\pi/\Omega$ is period of the classical trajectory,
 $\Omega \approx (3/2)\omega_{c}$.  The momentum $|p|$ is
$p = \sqrt{2m(V_{o}-E)} \approx \sqrt{2mV_{o}}$ since $E \sim \hbar \omega_{c}$
and $V_{o}>>E$. The dephasing time labelled ``linear" and ``quadratic" refer to
the cases in Sec.IV where the range of $[k]$ states covers the linear and
quadratic part of the spectrum, (fig.\protect\ref{figdeltav}).
We see from these estimates that eq.(\ref{osc}) can be satisfied when
$\tau_{in}>>10^{-11}$sec.

\vspace{0.2in}

{\bf Acknowledgement} : This work were completed during a very pleasant
sabbatical visit to HKUST and CUHK in Hong Kong, and an equally enjoyable
visit to NHMFL. I thank Nelson Cue, Kenneth Young, Hong-Ming Lai, and Bob
Schrieffer for their hospitality. This work is supported by NHMFL through
NSF Grant No. DMR 9016241.

\begin{table}
\caption{Parameters in Section IV}
\label{table1}
\begin{tabular}{ccc}
   &\multicolumn{2}{c}{$B=10$ Tesla} \\
   & \multicolumn{2}{c}{$L=1$cm} \\
   & $2a=100\AA$  &  $2a=60\AA$ \\
\tableline
$\omega_{c}$(sec$^{-1}$)  & $2.6\times10^{13}$ & {\rm same} \\
 (ev) & $1.7\times 10^{-2}$  &   {\rm same}\\
 ($^{o}$K)  & $200$  &   {\rm same}\\
$\ell$ ($\AA$)  & $81$  &  {\rm same}\\
$\Delta_{o}$(ev)  & $1.2\times 10^{-4}$  & $6.6\times 10^{-4}$\\
$\Delta_{o}$($^{o}K$)  & $1.4$  & $7.6$ \\
$\Delta_{o}/\hbar \omega_{c}$ & $7.1\times 10^{-3}$   &
$3.8\times 10^{-2}$ \\
$\tau^{}_{T}$(sec)  & $3.3 \times 10^{-11}$ & $6.3 \times 10^{-12}$\\
$\frac{\hbar v_{f}}{\Delta_{o}}\frac{2\pi}{L}$ & $7.1\times 10^{-4}$   &
$1.3\times 10^{-4}$ \\
linear $\tau^{(f)}_{dp}$(sec)    & $4.7\times 10^{-8}$ &
$4.7\times 10^{-8}$ \\
quadratic $\tau^{(f)}_{dp}$  & $1.3\times 10^{-4}$ &
$7.1\times 10^{-4}$ \\
\end{tabular}
\end{table}

\begin{figure}
\caption{A quantum Hall junction with a circular barrier. The trajectory
of a semiclassical electron is indicated by arrows. Once
tunnelled across the barrier, the electron will repeat a similar
reflected circular motion on the other side and eventually tunnel
back.}
\label{semiclassical}
\end{figure}

\begin{figure}
\caption{Schematic representation of the quantum
mechanical edge states near the barrier. The dotted lines labelled $a$
and $b$ denote two neighboring edge states, which in general have
different tunnelling rates across the barrier. As the flux $\phi$
through the hole at the center increases, the edge states will move
outward, thereby increasing their tunneling rates. State $b$ will evolve
to $a$ when $\phi=2\pi$.}
\label{ciredge}
\end{figure}

\begin{figure}
\caption{A rectangular version of the QH junction in
Figure~\protect\ref{semiclassical}. The dashed lines means that the
system is periodic in $y$ with period $L$.}
\label{rec}
\end{figure}

\begin{figure}
\caption{A quantum Hall weak link. The thick black lines represent
hard walls, i.e. infinite potentials. As we shall see, oscillatory
tunneling of the electrons across the weak link produces a singularity
in the noise of the tunneling current.}
\label{weaklink}
\end{figure}

\begin{figure}
\caption{A junction similar to Figure~\protect\ref{weaklink} except that both
$R$ an $L$ are multiply connected geometries. The current oscillation
across the junction will generate an oscillation of Hall voltage
across each ring.}
\label{recring}
\end{figure}

\begin{figure}
\caption{In the absence of external electric fields, a semiclassical electron
will be forever captured by the barrier once its trajectory is intercepted by
it.}
\label{capture}
\end{figure}

\begin{figure}
\caption{The spectrum of the rectangular QH system in figure~\protect\ref{rec}
in the
infinite barrier limit \protect\cite{Halperin}\protect\cite{MacDonald} :
Because of translational invariance in $y$, the sepctrum can be labelled
by the $y$-momentum $k$ and the Landua level index $n$.  The width of the
barrier is $2a$. Its boundaries are denoted in dimensionless units ($-a/\ell,
a/\ell$), where $\ell$ is the magnetic length.  The spectra of
$L$ and $R$, $\epsilon^{L}_{n,k}$ and $\epsilon^{R}_{n,k}$ are
represented by solid and dashed curves. They rise near
the barrier (i.e.  $|k|\rightarrow 0$) and intersect each
other. See also Section II for a detailed discussion. }
\label{inftyspec}
\end{figure}

\begin{figure}
\caption{The intersections of the spectra in figure \protect\ref{inftyspec}
turn into gaps as the infinite barrier becomes finite.  The
spectrum of the entire system will be denoted as $E_{n,k}$. They are
continuous curves that reduce to $\epsilon^{L}_{n,k}$ and
$\epsilon^{R}_{n,k}$ as one approaches the bulk, i.e. as $k<<-\ell^{-1}$
and $k>>\ell^{-1}$, where $\ell$ is the magnetic length.}
\label{general}
\end{figure}

\begin{figure}
\caption{A schematic representation of the wavefunctions of the
ground state $u_{0,k}(x)$ and first excited state $u_{1,k}(x)$ of
the entire system $L+R$ in figure~\protect\ref{general}.}
\label{figwf}
\end{figure}

\begin{figure}
\caption{The spectrum of the entire system in the presence of a voltage
bias. The lowest tunneling gap is now moved below the first bulk Landau level.}
\label{figbias}
\end{figure}

\begin{figure}
\caption{Both (a) and (b) show the sepctrum near the lowest tunneling gap
in the region $\ell k \leq 1$.
Part (a) is a schematic representation of the initial state $|\Psi>$ discussed
in Section IV when both $L$ and $R$ have identical chemical
potentials $\mu$.  Occupied (unoccupied) states are indicated by solid
(empty) circles. The spacing in $k$ is $2\pi/L$.  The states
$c^{}_{L,k}$ and $c^{}_{R,k}$ are linear combinations of $a^{}_{0,k}$
and $a^{}_{1,k}$. As the boundary condition in $y$ changes, (corresponding
threading a flux $\theta/2\pi$ through the hole in fig.~\protect\ref{ciredge}),
all $k$ states move to the right, i.e. 4 and 3 towards 3 and 2, 1' and 2'
towards 2' and 3', etc.
Part (b) shows the location of the $k$ states in
$(a)$ after half a flux is passes through the hole, i.e. $k\rightarrow
k+\pi/L$.
This results in
in an excess edge electron
on top of the Fermi surface in $L$.}
\label{before}
\end{figure}

\begin{figure}
\caption{When a chemical potential difference $eV$ is imposed between $R$
and $L$ at time $t=0$, the Fermi wavevectors in $L$ and $R$ is changed
to $-k_{F}+\delta k_{F}$ and $k_{F}+\delta k_{F}$. The range of $k$
contributing to the tunneling current is given by
$|k+k_{F}|<\delta k_{F}$, which is denoted as  $[k]$. In (a), the range
$[k]$ lies in the linear portion of the spectrum. The tunneling current
decays as $t^{-1}$. In (b), $[k]$ includes the intersection point, the
tunneling current decays as $t^{-1/2}$.  inside the gap.}
\label{figdeltav}
\end{figure}

\begin{figure}
\caption{Noise spectra at different chemical potentials:
When $\mu$ is below or above the gap, (a) and (c), only states in
the range $|k|>k_{F}$ contribute to the noise.  The noise spectrum has
a cusp.  When $\mu$ is inside the gap, all states
contribute to the current. The noise spectrum has a square root
singularity.}
\label{fignoise}
\end{figure}

\end{document}